\documentclass[aps,prl,superscriptaddress,reprint,showpacs,floatfix]{revtex4-1}
\usepackage{graphicx}
\usepackage{dcolumn}
\usepackage{bm}
\usepackage{xcolor}
\usepackage{relsize}
\usepackage{amsmath}
\usepackage{amsfonts}
\usepackage{mathtools}
\usepackage{braket}
\usepackage{gensymb}
\usepackage{url}
\usepackage{hyperref}
\usepackage{footmisc}
\usepackage[utf8]{inputenc}
\usepackage[euler]{textgreek}

\begin{document}

\preprint{APS/123-QED}

\title{Tunable photon statistics exploiting the Fano effect in a waveguide} 

\author{A.P. Foster}
\email{andrew.foster@sheffield.ac.uk}
\affiliation{Department of Physics and Astronomy, University of Sheffield, Sheffield S3 7RH, UK}
\author{D. Hallett}
\affiliation{Department of Physics and Astronomy, University of Sheffield, Sheffield S3 7RH, UK}
\author{I.V. Iorsh}
\affiliation{ITMO University, St. Petersburg 197101, Russia}
\author{S.J. Sheldon}
\affiliation{Department of Physics and Astronomy, University of Sheffield, Sheffield S3 7RH, UK}
\author{M.R. Godsland}
\affiliation{Department of Physics and Astronomy, University of Sheffield, Sheffield S3 7RH, UK}
\author{B. Royall}
\affiliation{Department of Physics and Astronomy, University of Sheffield, Sheffield S3 7RH, UK}
\author{E.~Clarke}
\affiliation{EPSRC National Epitaxy Facility, University of Sheffield, Sheffield S1 3JD, UK}
\author{I.A. Shelykh}
\affiliation{ITMO University, St. Petersburg 197101, Russia}
\affiliation{Science Institute, University of Iceland, Dunhagi 3, IS-107, Reykjavik, Iceland}
\author{A.M. Fox}
\affiliation{Department of Physics and Astronomy, University of Sheffield, Sheffield S3 7RH, UK}
\author{M.S. Skolnick}
\affiliation{Department of Physics and Astronomy, University of Sheffield, Sheffield S3 
7RH, UK}
\affiliation{ITMO University, St. Petersburg 197101, Russia}
\author{I.E. Itskevich}
\affiliation{Department of Engineering, University of Hull, Hull HU6 7RX, UK}
\author{L.R. Wilson}
\email{luke.wilson@sheffield.ac.uk}
\affiliation{Department of Physics and Astronomy, University of Sheffield, Sheffield S3 7RH, UK}

\date{\today}

\begin{abstract}

A strong optical nonlinearity arises when coherent light is scattered by a semiconductor quantum dot (QD) coupled to a nano-photonic waveguide. We exploit the Fano effect in such a waveguide to control the phase of the quantum interference underpinning the nonlinearity, experimentally demonstrating a tunable quantum optical filter which converts a coherent input state into either a bunched, or antibunched non-classical output state. We show theoretically that the generation of non-classical light is predicated on the formation of a two-photon bound state due to the interaction of the input coherent state with the QD. Our model demonstrates that the tunable photon statistics arise from the dependence of the sign of two-photon interference (either constructive or destructive) on the detuning of the input relative to the Fano resonance.

\end{abstract}

\maketitle

The generation of non-classical light is a fundamental requirement for the operation of quantum photonic devices. For instance, a single photon input is a prerequisite for linear optical quantum computation schemes \cite{RevModPhys.79.135}, while the use of NOON states may enable sensing with Heisenberg-limited precision in the field of quantum metrology \cite{PhysRevLett.118.257402,Bennette1501256}. On-demand, single-photon emitters such as quantum dots (QDs) are a proven resource for non-classical light, enabling the generation of single- \cite{Somaschi2016,PhysRevLett.116.020401} and two-photon \cite{Heindel2017} states, as well as the creation of entangled states on-chip \cite{Young2006}.

A markedly different approach to generate non-classical light (and to tune the photon statistics in general), involves the manipulation of a coherent input state, in such a way that the output state becomes either bunched or antibunched. A coherent input state can be considered as a weighted sum of different number-states \cite{Mark_Fox_2006}. Number-state filtering, in which the weighting of the individual number-states is controlled, can generate a quantum output state from such a classical coherent input (for instance via photon blockade \cite{Faraon2008,Reinhard2011,PhysRevLett.114.233601,PhysRevA.96.011801,PhysRevA.96.053810}). The use of interference has emerged as an extremely powerful tool in this regard: it has been shown theoretically that it can be used to realise complex photon statistics in cavity \cite{PhysRevA.83.021802,RADULASKI2017111_new,PhysRevLett.108.183601} and waveguide \cite{PhysRevApplied.7.044002,PhysRevA.97.023813} quantum electrodynamics (QED), and to generate single photons with simultaneous subnatural-linewidth using resonance fluorescence \cite{L_pez_Carre_o_2018}. Experimentally, the photon statistics of a coherent input have been manipulated via quantum interference in the weakly-coupled regime of cavity QED \cite{Bennett2016,DeSantis2017}, most recently using the unconventional photon blockade \cite{PhysRevLett.121.043601}. 

\begin{figure}[ht!]
\centering
\includegraphics[width=0.98\columnwidth]{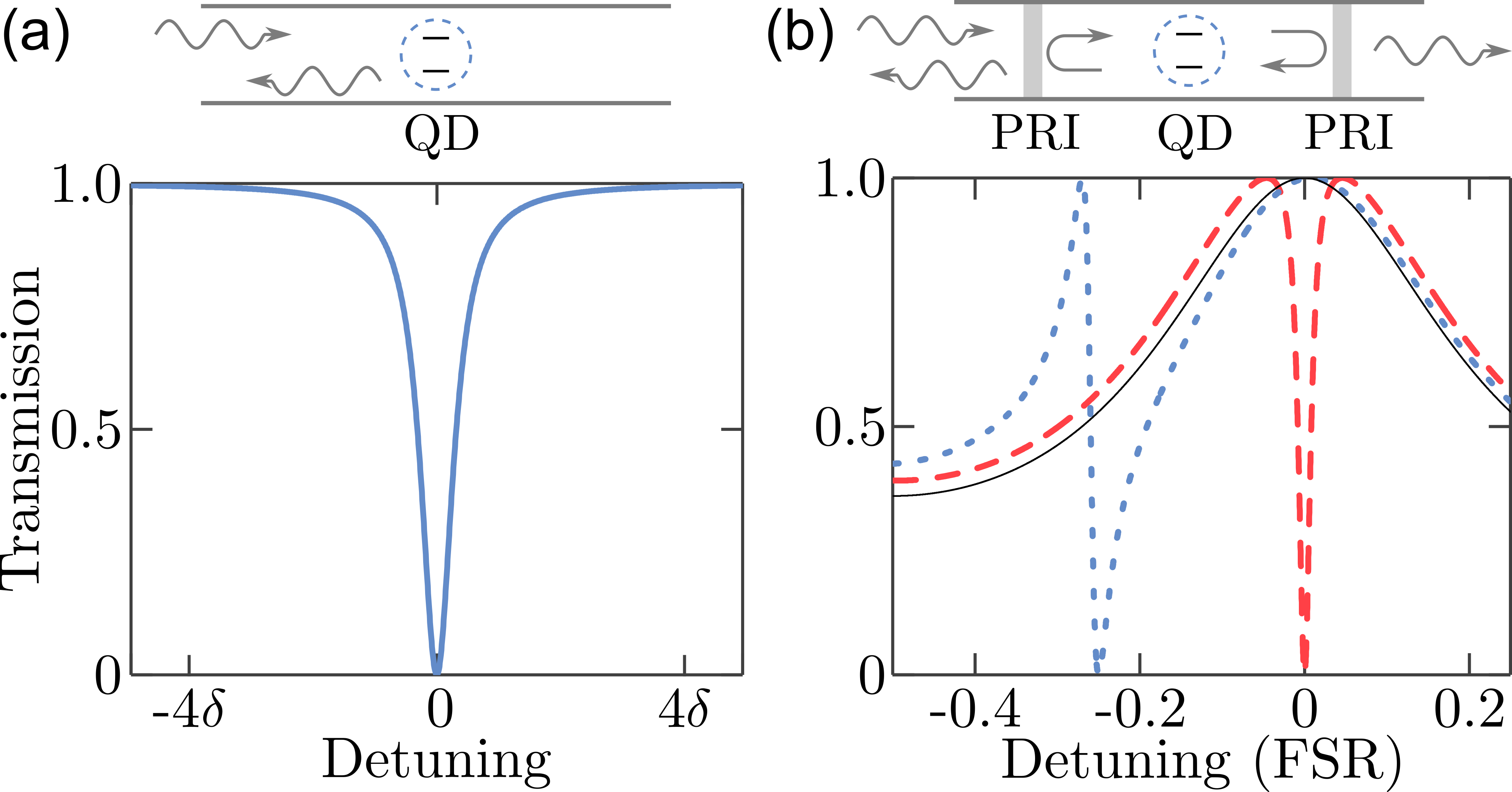}
\caption{\label{fig:One}(a) Calculated single photon transmission for a waveguide containing a single QD (upper schematic), as a function of the input detuning relative to the QD transition. $\delta$ is the QD transition linewidth. (b) Calculated transmission for a waveguide containing a single QD and supporting \mbox{F-P} modes due to partially reflective interfaces (PRIs) in the waveguide (upper schematic). The QD transition is either resonant (dashed red line) or non-resonant (dotted blue line) with the \mbox{F-P} mode. The transmission for the same waveguide without a QD is shown by a black solid line for reference. FSR stands for free spectral range.}
\end{figure}

An example of an interference phenomenon widely observed in photonics is the Fano effect \cite{Limonov2017}. A Fano resonance arises due to interference between a discrete transition and a background continuum, with the maxima and minima of the resulting spectral lineshape arising from constructive and destructive interference respectively. It has been shown theoretically that the detuning relative to the Fano resonance can be used to enable tunable number-state filtering \cite{PhysRevLett.105.263601,PhysRevA.94.043826,PhysRevApplied.7.044002}. To demonstrate this, we employ an integrated quantum photonic device comprising a single quantum two-level system, namely a QD, coupled to a single-mode optical waveguide. An ideal waveguide (with 100\% transmission) supports a background continuum of modes which have constant phase. Single photons resonant with the QD transition would be fully reflected due to destructive interference in the transmission direction between the continuum and photons scattered by the QD \cite{Chang2007}. This would result in a symmetric spectral profile in transmission, as shown in Fig.~\ref{fig:One}a. However, in a real device, reflections within the waveguide (see schematic in Fig.~\ref{fig:One}b) can lead to the formation of Fabry-P\'erot (\mbox{F-P}) modes, which modulate the transmission and hence the phase of the continuum. The spectral lineshape in transmission then depends on the detuning of the QD transition relative to the \mbox{F-P} modes, as shown in Fig.~\ref{fig:One}b. In particular, when the QD transition is detuned from a mode maximum, a characteristic Fano lineshape is observed \cite{Javadi2015,Hurst2018}.

For the output photon statistics of such a device, the behaviour of two-photon states is of importance. It has been predicted for an ideal waveguide that, on resonance with the discrete transition, two-photon states are preferentially transmitted (a manifestation of the nonlinear interaction between photons and the emitter at the single/few-photon level), and the output state is bunched \cite{Chang2007}. Similarly, bunching will occur when the input is detuned to the destructive interference regime of a Fano resonance. Notably however, when the input is detuned to the constructive interference regime, the possibility arises of the output state being antibunched \cite{PhysRevA.94.043826}. Such a non-classical output state, tunable across the Fano resonance, has yet to be demonstrated for an integrated quantum photonic device.

In this Letter, we demonstrate a tunable quantum optical filter using an integrated device comprising a single QD coupled to a single-mode nano-photonic waveguide. We inject a tunable, coherent laser field into the waveguide, and observe a Fano resonance in transmission. We show that the transmitted state photon statistics are antibunched when resonant with the Fano maximum and bunched at the Fano minimum, evidence of tunable number-state filtering. The tuning can be achieved either by changing the laser wavelength or by electrically Stark-shifting the QD transition, demonstrating control of the photon statistics locally, on-chip. We model the system and show that the formation of a two-photon bound (frequency entangled) state is critical to observe number-state filtering. Furthermore, antibunching is only achieved in the case of destructive interference of two-photon product states and bound states, which becomes possible due to the Fano resonance.  

\begin{figure}
\centering
\includegraphics[width=0.99\columnwidth]{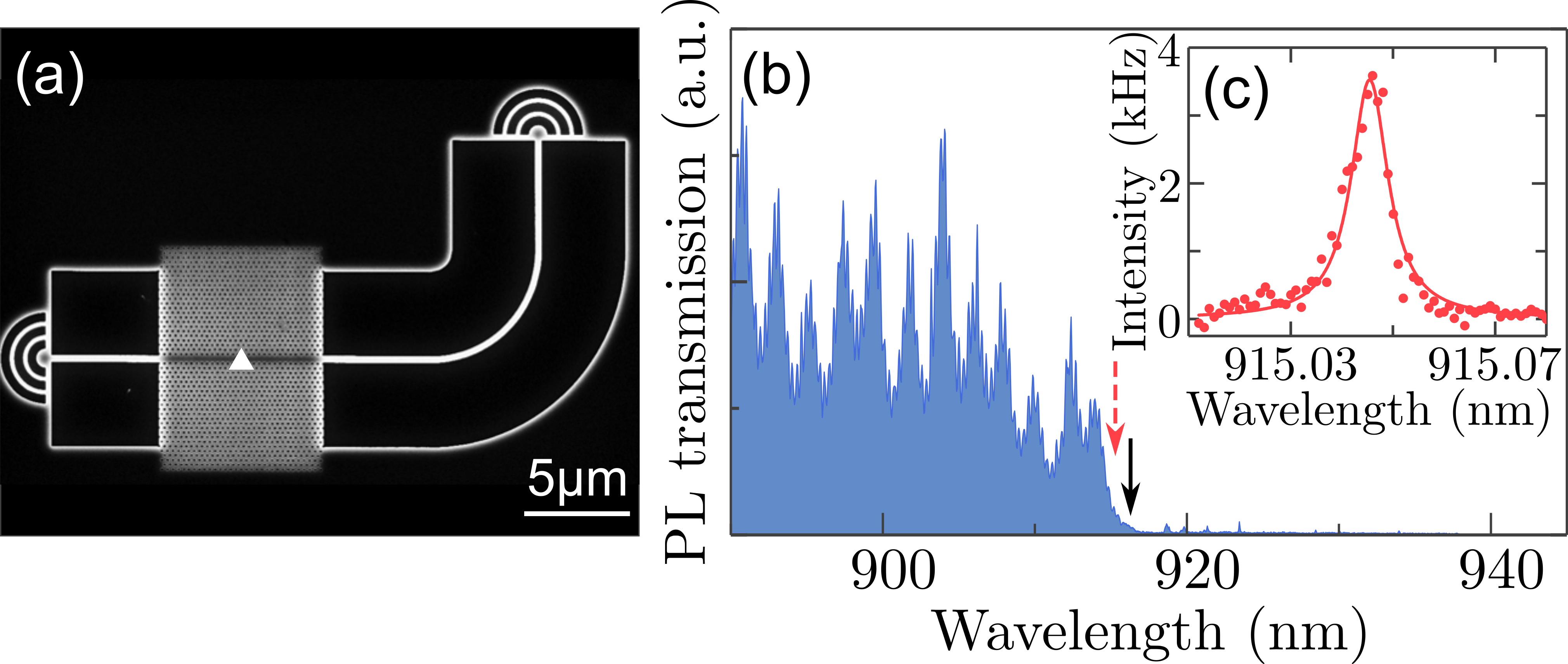}
\caption{\label{fig:Two}(a) Scanning electron microscope image of the nano-photonic device. The triangle shows the approximate location of the QD studied here, situated in a slow light PhCWG. (b) Device transmission probed using high power non-resonant PL (500\textmu W at 780nm). The black arrow indicates the location of the photonic band edge. (c) Resonant photoluminescence excitation spectrum for the trion state of the QD located in the PhCWG (circles), with spectral position given by the red dashed arrow in (b). The background laser scatter has been subtracted. The line is a Voigt fit to the data.}
\end{figure}

Fig.~\ref{fig:Two}a shows a scanning electron microscope image of our quantum optical filter, which was fabricated within a 170nm thick, GaAs \textit{p-i-n} membrane. InGaAs self-assembled QDs were embedded in the intrinsic region of the membrane and could be Stark tuned by application of a bias to the diode. (Details of the wafer, device design and experimental procedures can be found in the Supplemental Material \footnote{See Supplemental Material at for details concerning the sample design, experimental methods and additional results, and input-output modelling of the device. The Supplemental Material also contains an additional reference, Ref.~\cite{PhysRevA.85.023817}}.) The device consists of a suspended single-mode photonic crystal waveguide (PhCWG) with nanobeam waveguides attached to either end. The waveguides are terminated with semi-circular Bragg gratings which enable vertical in- and out-coupling of light. The PhCWG has a photonic band edge at $\sim$916nm, which was measured using waveguide-transmitted, non-resonant photoluminescence (PL) from the ensemble of QDs. The PL was excited in one Bragg coupler and detected from the other coupler, and is shown in Fig.~\ref{fig:Two}b. \mbox{F-P} modes  are revealed through oscillations in the transmitted intensity. The mode spacing of $\sim$2nm suggests that the dominant reflection occurs at the two Bragg coupler-nanobeam waveguide interfaces. 

QDs emitting in spectral proximity to the PhCWG band edge experience a slow light-induced Purcell enhancement \cite{Hughes2004}. The Purcell enhancement increases the QD exciton decay rate and consequently reduces the impact of dephasing on the coherence of the exciton emission. It also increases the $\beta$-factor which characterises the optical coupling strength between the QD and the waveguide mode \cite{PhysRevB.75.205437}, with a value as large as 0.98 previously reported \cite{PhysRevLett.113.093603}. In this regime, the QD may be considered as a ‘1D atom’, coupling almost uniquely to the single mode of the waveguide.

Resonance fluorescence measurements, with excitation from above the QD and collection from an outcoupler, were used to locate a suitable single QD in the PhCWG. Fig.~\ref{fig:Two}c shows the resonant photoluminescence excitation spectrum for such a QD, obtained by scanning a narrowband continuous wave laser across the QD transition. Its wavelength of 915.045nm lies within 1nm of the PhCWG band edge. The transition is likely to be a charged trion, as we typically observe fine-structure splitting for the QD neutral exciton in this sample \cite{Hallett2018}. In a separate measurement using resonant pulsed excitation (not shown), the lifetime of the trion state was found to be 150$\pm$30ps, which corresponds to a radiatively-limited linewidth of 4-6\textmu eV. We measured an ensemble lifetime of 750ps for QDs in the bulk of the sample and therefore estimate a Purcell factor of $\sim$5. The linewidth in Fig.~\ref{fig:Two}c is broadened to 15\textmu eV due to spectral wandering. We note that the QD could be Stark-tuned over more than 100\textmu eV, enabling full electrical control of the laser-trion \mbox{detuning}.

\begin{figure}
\centering
\includegraphics[width=1\columnwidth]{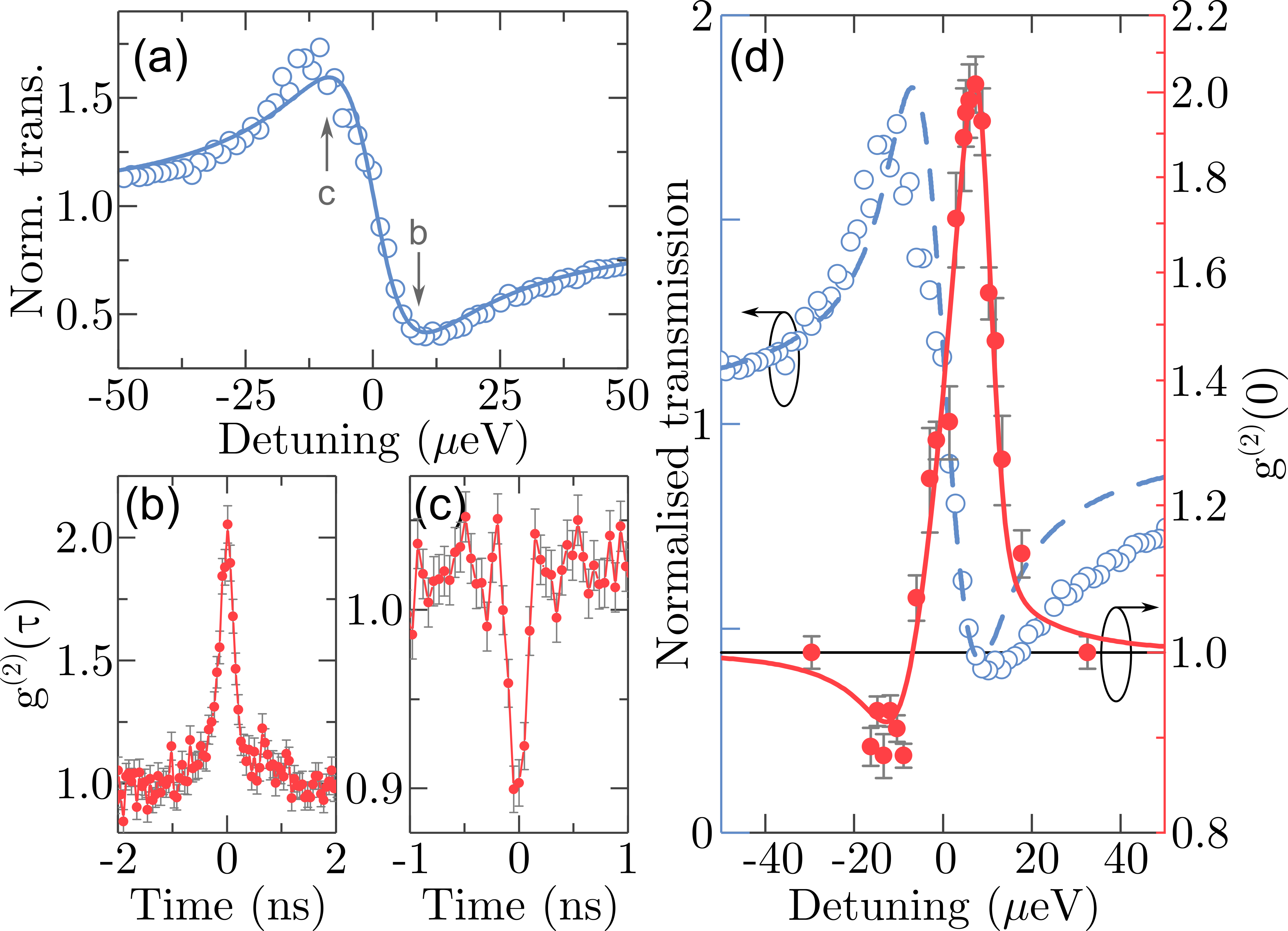}
\caption{\label{fig:Three} (a) Measured waveguide transmission as a function of laser-trion detuning (circles), normalised to the transmission measured at large detuning. The solid line is a Breit-Wigner-Fano fit to the data. (b,c) Second order autocorrelation function $g^{(2)}(\tau)$ at a detuning of (b) +9\textmu eV and (c) --9\textmu eV, as indicated in (a). The data has been normalised to the value of $g^{(2)}(\tau)$ at long time delay. Error bars correspond to the square root of the coincidence counts in each time bin. (d) $g^{(2)}(0)$ (red filled circles) and normalised waveguide transmission (blue open circles) as a function of laser-trion detuning. Error bars originate from fitting of the full $g^{(2)}(\tau)$ data. Solid and dashed lines represent the result of modelling (see text for details). The axis for $g^{(2)}(0)$ is logarithmic.}
\end{figure}

We next probe the effect of the same single QD on the waveguide transmission. A weak, tunable, continuous-wave laser was injected into the waveguide. The laser power was chosen such that, on average, less than one photon interacted with the QD within the trion lifetime. The transmission is therefore largely determined by the interaction of single photons with the QD. Fig.~\ref{fig:Three}a shows the transmission as a function of laser-trion detuning, which was controlled by changing the laser wavelength \footnote{It is also possible to control the detuning by Stark-shifting the energy of the QD transition, as shown in the Supplemental Material.}. The transmission is normalised to the background level measured in the absence of the laser-trion interaction. (This was achieved by electrical tuning of the QD transition far out of resonance with the laser.) A characteristic dispersive Fano lineshape is observed, due to interference between photons scattered from the QD and the driving laser field. Note that the minimum transmission is as small as 40\%, which is evidence for the strong interaction between single photons and the QD.

Now we consider the photon statistics of the transmitted field. Using a Hanbury Brown-Twiss setup, the second order autocorrelation function $g^{(2)}(\tau)$ was measured as a function of laser-trion detuning. The convolved instrument response time was 80ps. In Fig.~\ref{fig:Three}b-c we compare the normalised $g^{(2)}(\tau)$ histograms for laser-trion detunings of +9\textmu eV and \mbox{--9\textmu eV}, which correspond to the Fano transmission minimum and maximum respectively. At a detuning of +9\textmu eV substantial bunching is observed, with $g^{(2)}(0)$=2.02$\pm$0.07. After deconvolution with the instrument response function, we obtain a $g^{(2)}(0)$ value of 2.20$\pm$0.08, which is greater than the thermal classical limit. In sharp contrast, clear antibunching is measured at a detuning of --9\textmu eV, demonstrating the successful filtering out of two-photon states from the coherent input state. To demonstrate the tunability of our device, Fig.~\ref{fig:Three}d shows the measured $g^{(2)}(0)$ as a function of the laser-trion detuning, covering the full spectral width of the Fano resonance. From negative to positive detuning, a dispersive lineshape is seen in the photon statistics, whilst at large detuning the photon statistics are those of a coherent state, with $g^{(2)}(0)$ equal to unity. Thus, we have demonstrated that the photon statistics can be manipulated by means of number-state filtering using the detuning as the single control parameter. (See the Supplemental Material for measurements where the Stark shift of the QD transition was used as the control parameter.) 

Understanding of the output photon statistics requires consideration of two kinds of two-photon states, namely (separable) product states and (frequency entangled) bound states \cite{PhysRevA.76.062709,PhysRevLett.98.153003,PhysRevA.82.063816}. We note that for two-photon product states, constructive or destructive interference at the Fano resonance follows that of single-photon states. This implies that in the absence of bound states, a coherent input would always result in a coherent output. Observation of photon number-state filtering must therefore be related to the formation of the bound states. Indeed, the bound states have been shown to explain bunching \cite{PhysRevA.82.063816}. However, their presence alone is insufficient to explain the observed antibunching. To account for the antibunching, it is also necessary to consider interference between two-photon product states and bound states, as our following analysis shows. In particular, we identify the conditions under which antibunching becomes possible, and the role of \mbox{F-P} modes in this.

To gain the necessary insight, we model the system using the input-output formalism \cite{PhysRevA.94.043826}. The $g^{(1)}$ and $g^{(2)}$ two-time correlation functions are given by
\begin{align}
&g^{(1)}(t,t')=\frac{1}{t_0^2}\langle \alpha | \hat{c}_{out}^{\dagger}(t')\hat{c}_{out}(t)|\alpha\rangle,\\
&g^{(2)}(t,t')=\frac{\langle \alpha | \hat{c}^{\dagger}_{out}(t)\hat{c}^{\dagger}_{out}(t')\hat{c}_{out}(t')\hat{c}_{out}(t)|\alpha\rangle}{g^{(1)}(t,t)g^{(1)}(t',t')}\label{eq:g_2general},
\end{align}
where $t_0$ is the bare waveguide transmission amplitude in the absence of the QD, evaluated at the wavelength of the QD transition. The input coherent state is $\ket{\alpha}$, whilst $\hat{c}_{out}$ are the output field annihilation operators. In the low power, stationary limit ($t\rightarrow \infty$), with $\beta=1$ and neglecting QD dephasing, we find that

\begin{align}
g^{(1)}&=\frac{(\tilde{\delta}+\tan\phi)^2}{(1+\tilde{\delta}^2)}=|t_1|^2,\label{eq:g1}\\
g^{(2)}(0)&=\frac{1}{|t_1|^4}\left|t_1t_1+\frac{e^{2i\phi}}{T_0(1+\tilde{\delta}^2)}\right|^2 \label{eq:g2} \\
&=1+\frac{1}{T_0{^2}(\tilde{\delta}+\tan(\phi))^4}+\frac{2\cos2\phi}{T_0(\tilde{\delta}+\tan(\phi))^2}\label{eq:g2v2},
\end{align}
where $\tilde{\delta}$ is the detuning of the laser from the QD transition. The detuning is normalized to $\gamma/2$, where $\gamma$ denotes the transition decay rate. The single photon transmission amplitude is given by $t_1$, $T_0=t_0^2$ and $\phi=\mathrm{tan}^{-1}(-\sqrt{1-t_0^2}/t_0)$. Note that $\phi$ is related to the detuning of the QD transition from the \mbox{F-P} modes. The full time-dependent expressions, also accounting for QD dephasing and non-unity $\beta$-factor, can be found in the Supplemental Material. We note that in the limit of a weak pump, the expression for $g^{(2)}(0)$ can alternatively be derived by the scattering matrix (S-matrix) method~\cite{PhysRevA.94.043826,Fan2015}. Namely, in~\cite{PhysRevA.94.043826} an equation analogous to Eq.~\eqref{eq:g2} has been derived, where the first term comes from the disconnected S-matrix corresponding to the product state and the second term from the connected S-matrix corresponding to the bound two-photon state.

\begin{figure}
\centering
\includegraphics{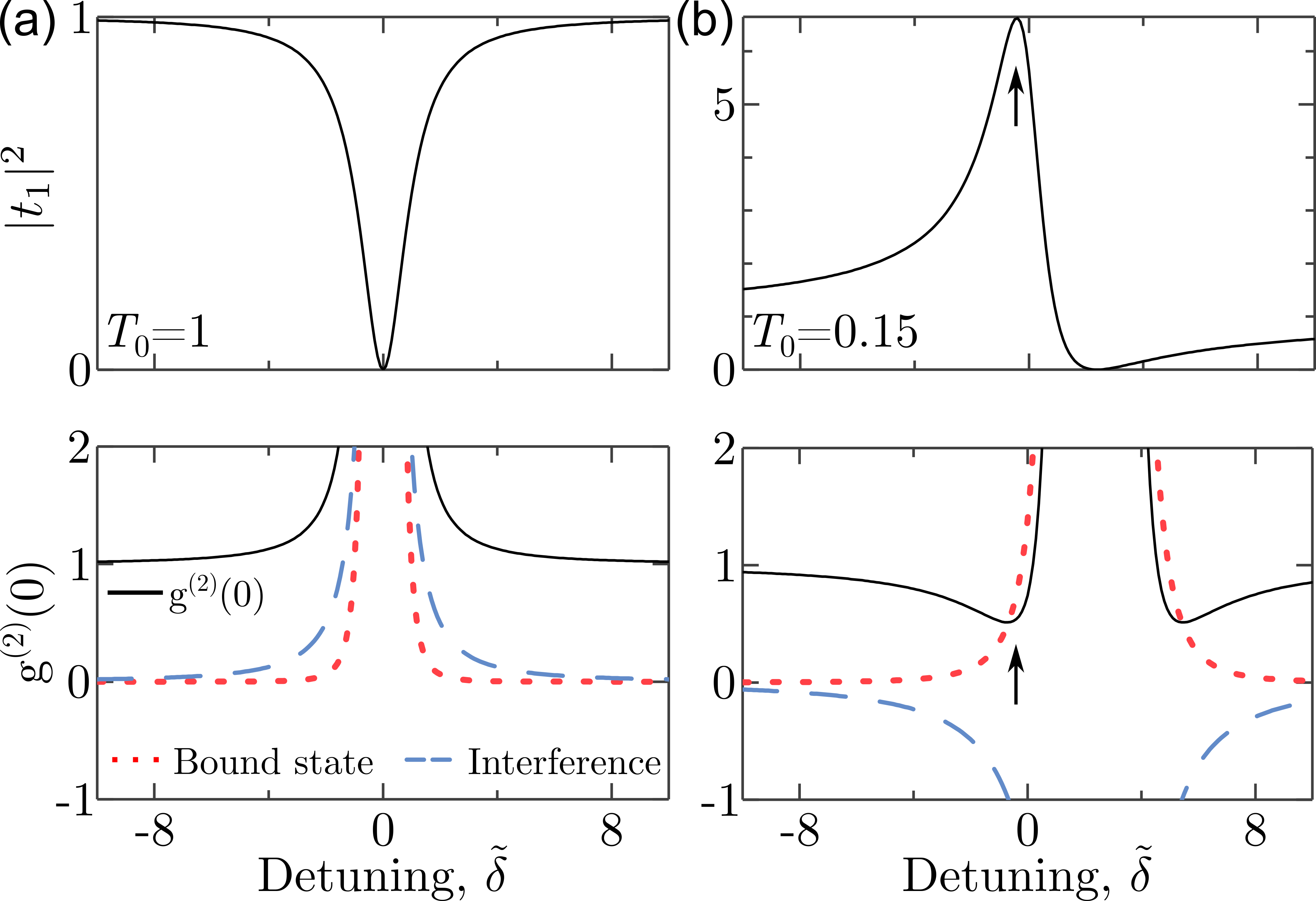}
\caption{\label{fig:Four} Theoretical transmission $|t_1|^2$ (upper panels) and $g^{(2)}(0)$ (lower panels - solid black lines) as a function of the detuning between the laser and the QD transition, for a bare waveguide transmission of (a) $T_0=1$ and (b) $T_0=0.15$. Antibunching can clearly be seen at a detuning corresponding to the Fano maximum, indicated by the arrows. Also shown are the contributions to the $g^{(2)}(0)$ from the two-photon bound state (red dotted lines) and the interference between two-photon product and bound states (blue dashed lines). The contribution to $g^{(2)}(0)$ from the two-photon product state is equal to unity for all detunings (see Eq.~\ref{eq:g2}).}
\end{figure}

We fit the model to the experimental $g^{(2)}(0)$ data, then use the same best-fit parameters to evaluate the normalised transmission using the equation for $g^{(1)}$. The resulting fits are shown in Fig.~\ref{fig:Three}d, showing very good agreement with the experimental results for both the transmission and the $g^{(2)}(0)$. The model clearly reproduces the most significant feature of the measured data, namely the generation of an antibunched transmitted field at negative detuning in addition to bunching at positive detuning. (See the Supplemental Material for more details of the fitting procedure.)

We now consider the physical process underpinning the quantum optical filter, using Eqs.~\eqref{eq:g2} and \eqref{eq:g2v2}. Eq.~\eqref{eq:g2} reveals interference between two-photon product states and two-photon bound states \cite{PhysRevA.94.043826} (the first and the second term in the modulus squared, respectively). The first two terms in Eq.~\eqref{eq:g2v2}, obtained after evaluation of the modulus, represent the bare contributions to $g^{(2)}(0)$ from the product states (the unity term) and the bound states, respectively, and the third term describes interference between the two-photon states. Analysis of Eqs.~\eqref{eq:g2} and \eqref{eq:g2v2} leads to several immediate conclusions. First, it is clear that the formation of the bound state is critical for the generation of non-classical light, as in its absence $g^{(2)}(0)$ equals unity. Secondly, the bound state term in Eq.~\ref{eq:g2v2} is positively valued, and the first two terms combined would only lead to $g^{(2)}(0)\geq1$, i.e. either a coherent or bunched output state. Evidently, antibunching can only arise in the case of destructive two-photon interference, for which the third term in Eq.~\ref{eq:g2v2} is negative. The latter is possible in the case when $\phi<-\pi/4$ and hence for bare waveguide transmission $T_0<0.5$. One should note that access to this regime in an otherwise ideal waveguide is enabled by the presence of \mbox{F-P} modes (whose presence also gives rise to the Fano effect.) 

This is illustrated in Fig.~\ref{fig:Four}, in which we plot the transmission $|t_{1}|^{2}=g^{(1)}$ and $g^{(2)}(0)$, for representative values of $T_0=1$ and $T_0=0.15$. $T_0=1$ corresponds to a QD in a perfectly transmissive waveguide, while at $T_0=0.15$ a QD is significantly off-resonant with an \mbox{F-P} mode. When $T_0=1$, both the bound state contribution and the interference term are positive, and bunching is predicted across the whole range of detuning in Fig.~\ref{fig:Four}a, in agreement with Ref.~\cite{PhysRevA.94.043826}. However, for $T_0=0.15$ the interference term is negative, and where it outweighs the contribution from the bound state, antibunching occurs. Notably, one expects antibunching at the Fano transmission maximum, as observed experimentally. (This is indicated by arrows in Fig.~\ref{fig:Four}b). In particular, for $T_0=0.15$, $g^{(2)}(0)$ is expected to be as low as 0.5 at the Fano maximum; furthermore, $g^{(2)}(0)\rightarrow0$ as $T_0\rightarrow0$ in the ideal case scenario. Physically, the Fano maximum favours transmission of single photons from the coherent input. At the same time, the contribution to the output from the two-photon product states is suppressed due to destructive interference with the bound states, resulting in antibunching.

The tunable number-state filtering effect, which we observe, is therefore critically dependent on two factors: the formation of a two-photon bound state, which enables non-classical light to be generated in the first instance; and the destructive interference of the two-photon product state and bound state.  The filter switches the output between bunched and antibunched, dependent on the strength of the destructive interference effect; this in turn depends on the detuning relative to the Fano resonance.

In conclusion, we have demonstrated an integrated, tunable quantum optical filter which exploits the Fano effect to convert a coherent input state into either a bunched, or antibunched non-classical output state, and provided its theoretical analysis.  The filter is formed from a single QD coupled to a nano-photonic waveguide, which supports Fabry-P\'erot modes. A Fano resonance is observed in the waveguide transmission as a coherent input laser is tuned across the QD transition. Antibunching of the output state is observed when the laser is resonant with the Fano maximum, and bunching at the Fano minimum. Switching between the two states is achieved by controlling the detuning of the laser relative to the Fano resonance, either by changing the laser wavelength, or locally, using the quantum-confined Stark effect. Notably, antibunching is only observed due to the presence of the Fano resonance. We have shown theoretically that the non-classical output state is critically dependent on the formation of a two-photon bound state due to interaction of the coherent input with the QD, and that control over the photon statistics arises due to the change between constructive and destructive two-photon interference at the extrema of the Fano resonance. Our results offer a new direction for the use of quantum interference effects in integrated photonic circuits; in particular, the Fano resonance is of significant interest for applications requiring fast optical switching \cite{Limonov2017,doi:10.1063/1.4893451} and our work extend this capability to switching of photon statistics.

Data supporting this study are openly available from the University of Sheffield repository \footnote{\MakeLowercase{h}ttps://doi.org/10.15131/shef.data.7360988}.

\begin{acknowledgments}
This work was supported by EPSRC Grant No. EP/N031776/1 and by Megagrant 14.Y26.31.0015 of the Russian Federation. I.V.I. thanks D. Kornovan for valuable discussions. 
\end{acknowledgments}

\bibliography{References} 
\clearpage
\pagebreak

\renewcommand{\theequation}{S\arabic{equation}}
\renewcommand{\thefigure}{S\arabic{figure}}
\setcounter{figure}{0}
\setcounter{equation}{0}
\onecolumngrid

\section*{Supplemental Material for \\ ``Tunable photon statistics exploiting the Fano effect in a waveguide''}

\section{Sample design and experimental details}
\subsection{Wafer structure}
\begin{figure}[h]
\centering
\includegraphics[scale=1.5]{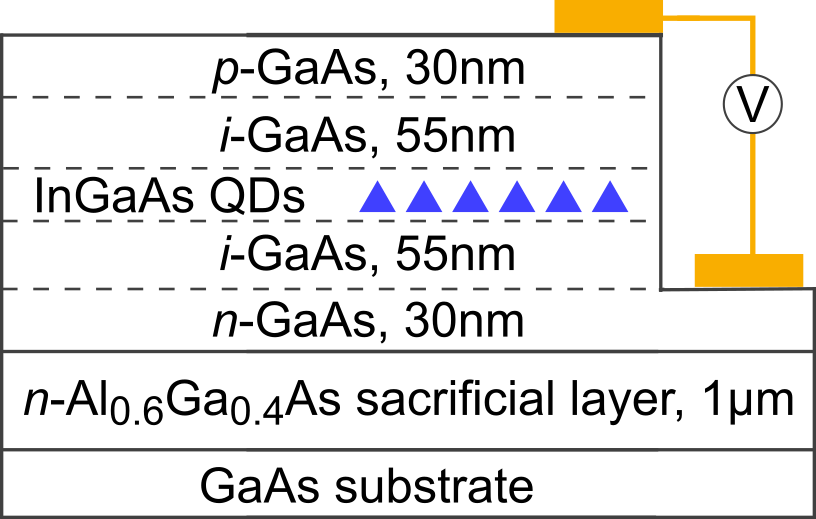}
\caption{\label{fig:S1} Schematic of the epitaxial layer structure.}
\end{figure}

The epitaxial layer structure for our quantum optical filter is shown in Fig.~\ref{fig:S1}. The 170nm thick \textit{p-i-n} membrane contains a layer of \mbox{InGaAs} self-assembled QDs emitting between 880nm and 940nm. To fabricate the device, an 80nm thick Si0$_{\text{x}}$ hard mask was first deposited on the wafer. The nano-photonic structure was then defined in the   membrane using electron beam lithography and reactive ion etching, after which the hard mask was removed using a 1\% hydrofluoric acid wet etch.  Ti/Au (20nm/200nm) electrical contacts were made to the \textit{p} and \textit{n} layers of the GaAs membrane. Finally, the AlGaAs sacrificial layer was removed from beneath the structure using a 10\% hydrofluoric acid wet etch.

\vspace{10pt}

\subsection{Waveguide design}

For a complete description of the photonic crystal waveguide design see the Supporting Materials for Ref. \cite{Hallett2018}. 
Note that the 90 degree bend in one of the nanobeam waveguides (see Fig. 2a in the main text) allowed for the use of a cross-polarised excitation and collection geometry, to minimise detection of undesirable laser scatter during resonance fluorescence and resonant transmission experiments \cite{Javadi2015,Hallett2018}. 

\subsection{Power dependence of the waveguide transmission}

The transmission measurements in the main text were taken using a power of 1\textmu W, just below the onset of saturation of the QD. The power was measured before the objective lens above the sample, and the fraction of this power coupled into the waveguide is dependent on the Bragg grating coupler efficiency. By comparison with Ref. \cite{Javadi2015} we conclude that on average, less than one photon was incident on the QD per trion lifetime.

\subsection{Hanbury Brown and Twiss (HBT) measurements}
For HBT measurements, the transmitted light was split using a 50:50 fibre beam splitter, and detected using two superconducting nanowire single-photon detectors (Single Quantum). Correlations between detection events were evaluated using a time correlated single-photon counting card (Becker-Hickl SPC-130). The instrument response time  for the measurement was 80ps (full width half maximum).

\subsection{Detuning-dependent HBT using the QD Stark shift}
\begin{figure}[h]
\centering
\includegraphics[width=0.5\columnwidth]{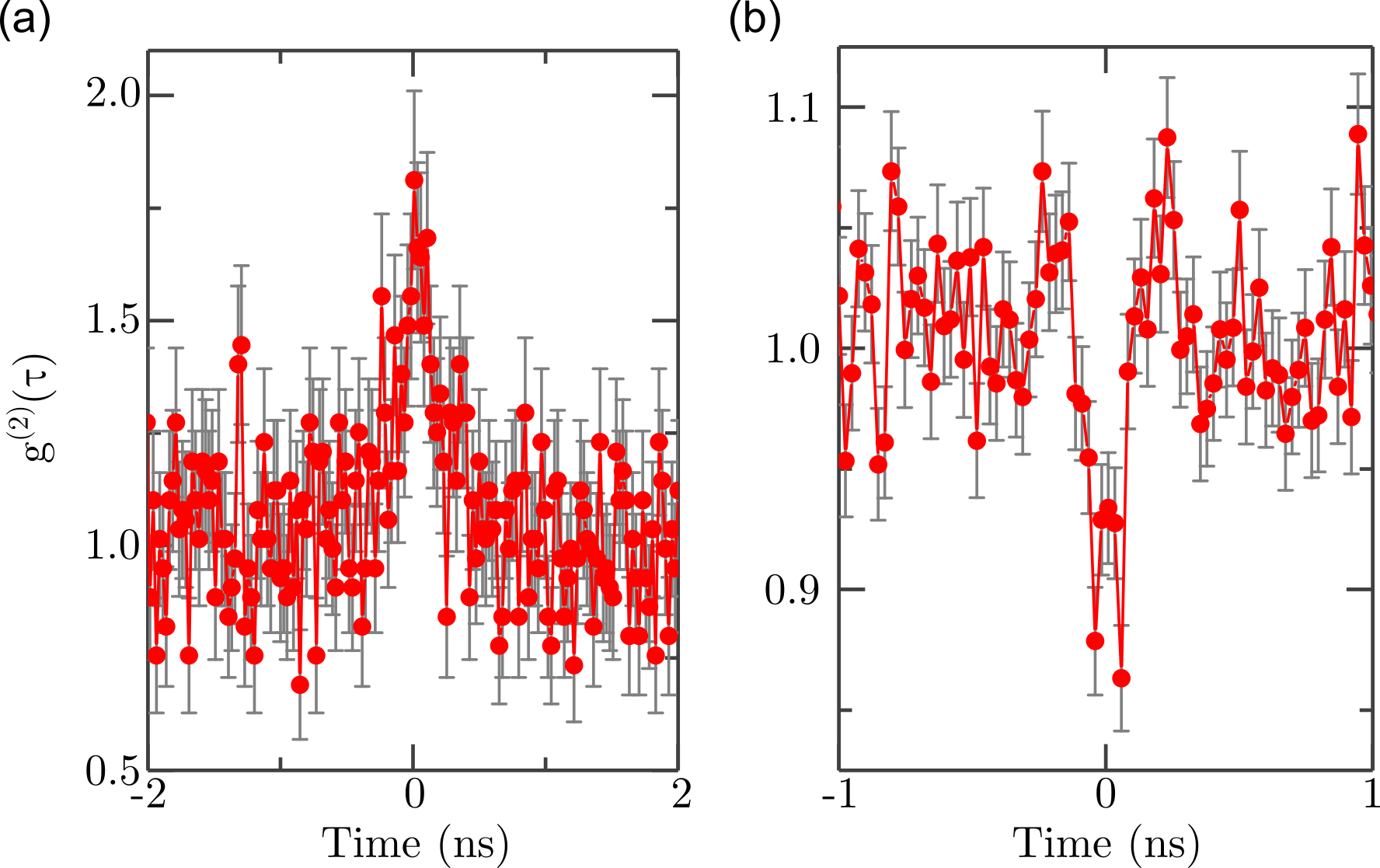}
\caption{\label{fig:S2} Second order autocorrelation function $g^{(2)}(\tau)$ at a detuning of (a) +9\textmu eV and (b) --9\textmu eV. The data has been normalised to the value of $g^{(2)}(\tau)$ at long time delay. Error bars correspond to the square root of the coincidence counts in each time bin.}
\end{figure}

The detuning-dependent HBT measurements reported in the main text were undertaken using the laser wavelength as the control parameter, with a fixed bias applied to the \textit{p-i-n} diode. It is also possible to locally control the detuning via the Stark shift, whilst keeping the laser wavelength fixed. In Fig.~\ref{fig:S2} we show HBT measurements undertaken by changing the diode bias. Bunching (antibunching) is seen at a detuning of +9\textmu eV (--9\textmu eV), in agreement with the measurements in the main text. The measurement noise is larger here than for the equivalent measurements in the main text, due to a shorter data collection time.

\vspace{10pt}

\section{Input-output model}
 \subsection{Relation between coupling coefficient, bare waveguide transmission and QD lifetime}
We start with the equations of motion for the QD operator supplemented with the relation between the ingoing and outgoing amplitude (see e.g. Ref.~\cite{PhysRevA.94.043826})
 \begin{align}
&\frac{d}{dt}\hat\sigma_{-}=-i\omega_0\hat\sigma_--(\gamma/2+\gamma_{de})\hat\sigma_--\hat\sigma_z \mathbf{d}^{T}\mathbf{c_{in}},\\
&\frac{d}{dt}\hat\sigma_z=-\gamma\hat\sigma_z-\gamma +2\hat\sigma_{+}\mathbf{d}^{T}\mathbf{c}_{in}+2\mathbf{c_{in}^{\dagger}}d^*\hat\sigma_-,\\
&\mathbf{c_{out}}(t)=\mathbf{C}\mathbf{c_{in}}+\hat\sigma_-(t)\mathbf{d},\label{eq:out-in}
\end{align}
where $\omega_0$ denotes the QD transition frequency. The radiative and pure dephasing decay rates for the transition are $\gamma$ and $\gamma_{de}$, respectively. The vectors $\mathbf{c}_{in}$ and $\mathbf{c}_{out}$ correspond to the ingoing and outgoing waves respectively in the left/right propagation basis, $\mathbf{C}$ is the scattering matrix of the whole structure in the absence of the QD \cite{PhysRevA.94.043826}, $\mathbf{d}$ is the coupling of the QD to the left and right moving waves, and $\hat{\sigma}_{l}(t)$ with $l=\{\pm,z\}$ are the QD raising and lowering operators. In order to determine the connection between the experimentally measured lifetime $1/\gamma$ and the coupling coefficient vector $\mathbf{d}$, we employ the flux conservation condition. Specifically, we write the continuity equation
\begin{align}
\frac{d}{dt}N=\mathbf{c_{in}^{\dagger}c_{in}}-\mathbf{c_{out}^{\dagger}c_{out}}-(1-\beta)\gamma N,\label{dNdt}
\end{align}
where $N=\hat\sigma_+\hat\sigma_-=(\hat\sigma_z+1)/2$ is the population operator for the QD. Eq.~\eqref{dNdt} states that in equilibrium the change in the population of the QD is proportional to the difference between the ingoing and outgoing energy flux. The final term describes the energy decay channel corresponding to non-radiative decay of the QD or radiative decay into modes other than the single mode of the waveguide (i.e. to the farfield). The parameter $\beta$ is the probability of the excited QD eventually decaying via emission of a photon into the waveguide mode (referred to as the $\beta$-factor in the main text). 

At the same time, we can write down another equation for $N$ from the equation for $\sigma_z$, which reads
\begin{align}
\frac{d}{dt}N=-\gamma N +\hat\sigma_{+}\mathbf{d}^{T}\mathbf{c}_{in}+\mathbf{c_{in}^{\dagger}}d^*\hat\sigma_-.
\end{align}

Substituting in Eq.~\eqref{eq:out-in} for the outgoing amplitudes we obtain the equation:
\begin{align}
0=\mathbf{c_{in}^{\dagger}}[\mathbf{I}-\mathbf{C^{\dagger}C}]\mathbf{c_{in}}-(\mathbf{d^{\dagger}d}-\beta\gamma)N-\mathbf{c_{in}^{\dagger}}[\mathbf{C^{\dagger}d+d^*}]\hat\sigma_--\hat\sigma_+[\mathbf{d^{\dagger}C}+\mathbf{d}^T]\mathbf{c_{in}}.
\end{align}
We then make the assumption that our non-resonant scattering matrix is a unitary one, hence $\mathbf{C^{\dagger}C}=\mathbf{I}$, i.e. we have no scattering losses in the empty waveguide. 
$\mathbf{C}$ can now be represented in the form
\begin{align}
\mathbf{C}=\begin{pmatrix}
t_0 & ir_0 \\ ir_0 & t_0
\end{pmatrix},
\end{align}
where $t_0\in [0,1]$ is the transmission of the structure in the absence of the QD, and $r_0=-\sqrt{1-t_0^2}$ is the reflection coefficient. The relations between $\mathbf{C},\mathbf{d}$ and $\gamma$ are given by
\begin{align}
\mathbf{d^{\dagger}d}=\beta\gamma,\\
\mathbf{C^{\dagger}d=-d^*},
\end{align}
therefore the coupling $\mathbf{d}$ can be uniquely defined as
\begin{align}\label{eq:d_expr}
\mathbf{d}=ie^{i\phi/2}\sqrt{\beta\gamma/2}\begin{pmatrix}
1 \\ 1
\end{pmatrix},
\end{align}
where \mbox{$\phi=\arctan(r_0/t_0)$}.
The value of $\gamma$ can be directly measured experimentally. It should be noted that $\gamma$ is renormalized due to the cavity-induced Purcell effect. At the same time, $r_0$ can be only roughly estimated and is a fitting parameter in the simulations. In deriving expression~\eqref{eq:d_expr} we have implicitly assumed that the QD is placed in the cavity centre.

\subsection{Calculation of $g^{(1)}$ and $g^{(2)}$ correlation functions}
The two-time correlation functions $g^{(1)}$ and $g^{(2)}$ by definition are
\begin{align}
&g^{(1)}(t,t')=\frac{1}{t_0^2}\langle \alpha | \hat{c}_{out}^{\dagger}(t')\hat{c}_{out}(t)|\alpha\rangle,\label{eq:g_1general}\\
&g^{(2)}(t,t')=\frac{\langle \alpha | \hat{c}^{\dagger}_{out}(t)\hat{c}^{\dagger}_{out}(t')\hat{c}_{out}(t')\hat{c}_{out}(t)|\alpha\rangle}{g^{(1)}(t,t)g^{(1)}(t',t')}\label{eq:g_2general},
\end{align}
where $\ket{\alpha}$ is the input coherent state and $\hat{c}_{out}$ are the output field annihilation operators.

In the following equations, $\hat{c}_{in}$ are the input field annihilation operators,  $\omega$ is the input photon frequency and $\alpha$ is the coherent field amplitude. The output annihilation operators obey the relationship
\begin{align}
\hat{c}_{out}=t_0\hat{c}_{in}(t)+\hat{\sigma}_{-}(t)d,
\end{align}
where $\hat{c}_{in}(t)$ is defined as
\begin{align}
\label{eq:c_in}
\hat{c}_{in}(t)|\alpha\rangle=\alpha e^{-i\omega t}|\alpha\rangle.
\end{align}
The output field function in Eq.~\eqref{eq:g_2general} can thus be expressed via the correlators of the QD raising and lowering operators. These obey the set of differential equations
\begin{align}
&\frac{d}{dt}{\hat{\sigma}}_{-}=-i\omega_0\hat{\sigma}_{-}-(\frac{\gamma}{2}+\gamma_{de})\hat{\sigma}_{-}-\hat{\sigma}_zd\alpha e^{-i\omega t},\label{eq:sys1}\\
&\frac{d}{dt}{\hat{\sigma}}_{+}=i\omega_0\hat{\sigma}_{+}-(\frac{\gamma}{2}+\gamma_{de})\hat{\sigma}_{+}-\hat{\sigma}_z d^*\alpha^* e^{i\omega t},\label{eq:sys2}\\
&\frac{d}{dt}{\hat{\sigma}}_{z}=2(d\alpha e^{-i\omega t}\hat{\sigma}_{+}+d^*\alpha^* e^{-\omega t}\hat{\sigma}_{-})-\gamma\hat{\sigma}_{z}-\gamma.\label{eq:sys3}
\end{align}
We seek solutions for the operator averages $\langle \hat{\sigma}_{l}(t)\rangle$, $l=\{\pm,z\}$ and for the correlators $\langle \hat{\sigma}_{+}(t)\hat{\sigma}_{l}(t+\tau)\rangle$, and $\langle \hat{\sigma}_{+}(t)\hat{\sigma}_{l}(t+\tau)\hat{\sigma}_{-}(t)\rangle$ which enter the expressions for the $g^{(2)}$ function. These correlators are found by using the quantum regression theorem \cite{PhysRevA.85.023817}. Namely, for the $g^{(1)}$ correlation function in the limit $t=t'\rightarrow\infty$ we obtain
\begin{align}
g^{(1)}=1-\frac{2\beta(\zeta-\delta\tan\phi)}{\tilde{\alpha}\zeta+\zeta^2+\delta^2}+\frac{\beta^2}{\cos^2\phi(\zeta\tilde{\alpha}+\zeta^2+\delta^2)},\label{g_1ans}
\end{align}
where $\zeta=1+2\gamma_{de}/\gamma$ is the dimensionless measure of the QD decoherence, $\tilde{\delta}=2(\omega-\omega_0)/\gamma$ is the dimensionless detuning, and $\tilde{\alpha}=4\beta|\alpha|^2/\gamma$ is the dimensionless intensity of the coherent input state. In the limit $\beta=1$, $\zeta=1$ and vanishing intensity,  Eq.~\eqref{g_1ans} coincides with the single photon transmission probability derived in Ref.~\cite{PhysRevA.94.043826} (Eq.[49]):
\begin{align}
\beta=1, \zeta=1:\quad g^{(1)}=\frac{(\tilde{\delta}+\tan\phi)^2}{\tilde{\delta}^2+1}.
\end{align}
The expression for the $g^{(2)}$ correlation function in the limit of a weak pump is given by
\begin{align}
g^{(2)}(\tau)=\frac{1}{(g^{(1)})^2}\left[(g^{(1)})^2+Ae^{-2|\tau|}+(g^{(2)}(0)-(g^{(1)})^2-A)\cos(\tilde{\delta}\tau)e^{-\zeta|\tau|}+B\sin(\tilde{\delta}|\tau|)e^{-\zeta |\tau|}\right],\label{g_2}
\end{align}
where the dimensionless $\tau$ is normalized to $\gamma/2$, $g^{(2)}(0)=g^{(1)}|_{\beta=2\beta}$,
\begin{align}
A=\frac{\beta^3(\beta-4(\zeta-1)\cos^2\phi)}{\cos^4\phi (\zeta^2+\tilde{\delta}^2)(\tilde{\delta}^2+(\zeta-2)^2)},
\end{align}
and

\begin{align}
&B=
2\beta^4\frac{(1-\zeta)(\zeta^2-2\zeta+\tilde{\delta}^2)}{\tilde{\delta}(\zeta^2+\tilde{\delta}^2)^2(\tilde{\delta}^2+(\zeta-2)^2)\cos^4\phi}
+2\beta^2\frac{2\zeta\tilde{\delta}\cos2\phi-(\tilde{\delta}^2-\zeta^2)\sin 2\phi}{\cos^2\phi(\zeta^2+\tilde{\delta}^2)^2}\nonumber\\
&-4\beta^3\frac{\cos\phi\left[\tilde{\delta}^2(\tilde{\delta}^2+2-\zeta)-\zeta^2(\zeta-2)(\zeta-1)\right]+\sin\phi\left[\tilde{\delta}\zeta(\tilde{\delta}^2+(\zeta-2)^2)\right]}{\tilde{\delta}(\zeta^2+\tilde{\delta}^2)^2(\tilde{\delta}^2+(\zeta-2)^2)\cos^3\phi}.
\end{align}
Note that the decoherence affects the interference terms, as $\zeta$ enters the argument of the exponent for these terms only. In the limit $\beta=1$ and $\zeta=1$, Eq.~\eqref{g_2} coincides up to a constant prefactor with Eq.[51] in Ref.~\cite{PhysRevA.94.043826}. 
In this case we can write the expression for $g^{(2)}$ as
\begin{align}
g^{(2)}(\tau)=\frac{1}{(g^{(1)})^2}\left|g^{(1)}+\left(\frac{e^{-|\tau|}}{t_0^2(\tilde{\delta}^2+1)}\right)e^{i(2\phi+\tilde{\delta} |\tau|)}\right|^2.
\end{align}
At zero delay time we obtain 
\begin{align}
g^{(2)}(0) = 1+\frac{1}{t_0^4(\tilde{\delta}+\tan\phi)^4}+\frac{2\cos 2\phi}{t_0^2(\tilde{\delta}+\tan\phi)^2},
\end{align}
{which is Eq. 5 within the main text.}

\section{Fitting procedure}
In order to provide a fit for the experimental data, both $g^{(1)}$ and $g^{(2)}$ in Eqs.~\eqref{g_1ans},~\eqref{g_2} were first averaged over the QD transition energy, which varies due to spectral wandering. We assume a Gaussian distribution for the spectral wandering with standard deviation $\sigma$. Then, introducing the dimensionless $\tilde{\sigma}=2\sigma/\gamma$, we write the averaged $g^{(1)}$ as
\begin{align}
\langle g^{(1)}\rangle(\tilde\delta)=\frac{1}{\sqrt{2\pi}\tilde{\sigma}} \int dx g^{(1)}(\tilde{\delta}+x)e^{-x^2/2\tilde{\sigma}^2},\label{g1vaerall}
\end{align}
where $\langle\rangle$ denotes averaging over spectral wandering. In averaging the $g^{(2)}$ correlation function we first average separately the numerator and denominator, and then divide:
\begin{small}
\begin{align}
\langle g^{(2)}\rangle(\tau,\delta)=\frac{\langle\left[(g^{(1)})^2+Ae^{-2|\tau|}+(g^{(2)}(0)-(g^{(1)})^2-A)\cos(\tilde{\delta}\tau)e^{-\zeta |\tau|}+B\sin(\tilde{\delta}|\tau|)e^{-\zeta |\tau|}\right]\rangle}{\langle (g^{(1)})^2\rangle}.
\end{align}
\end{small}

The spectral averaging smears the sharp features in the theoretical transmission and $g^{(2)}$. For example, for $\beta=1$ and in the absence of spectral wandering, arbitrarily small normalized transmission and thus $g^{(1)}$ is predicted, with a consequently huge increase in $g^{(2)}$. However, if spectral wandering is present the minimal transmission is limited by $\tilde{\sigma}/(\tilde{\sigma}+1)$.

Finally, the finite response time of the detectors should be taken into account. The response function of the detectors is a Gaussian with a standard deviation of $t_{resp}=34$ps (80ps FWHM). Introducing the dimensionless $\tilde{t}_{resp}=2t_{resp}/\gamma$, we write down for the temporally convolved $g^{(2)}$

\begin{align}
\langle\langle g^{(2)}(\tilde\delta,\tau)\rangle\rangle=\frac{1}{\sqrt{2\pi}\tilde{t}_{resp}}\int d\tau' \langle g^{(2)}(\tilde\delta,\tau+\tau')\rangle e^{-\tau'^2/2\tilde{t}_{resp}^2}  \label{g2averall}
\end{align}
where $\langle\langle\rangle\rangle$ denotes averaging over time after averaging over spectral wandering. We fit the experimental data using Eqs.~\eqref{g1vaerall} and \eqref{g2averall}, with fitting parameters $\omega_0,\gamma,\gamma_{de},\sigma,t_0$ and $\beta$. We find that $\omega_0=327.524$THz, $1/\gamma=125$ps, $1/\gamma_{de}=38$ns, $\sigma=4.7$\textmu eV, $t_0=0.62$ and $\beta=0.99$. Reasonable agreement is found between the fitted values and those parameters which were measured independently (the lifetime (150$\pm$30ps) and degree of spectral wandering (standard deviation of 5.3\textmu eV)).
\bibliography{References}

\end{document}